\documentclass[a4paper,10pt]{article}

\usepackage{ucs}
\usepackage[utf8x]{inputenc}

\usepackage{a4wide,url}
\usepackage[pdftex]{graphicx}
\usepackage[colorlinks,hypertexnames=false,pdftex]{hyperref}
\usepackage{tabulary}
\usepackage{booktabs}
\usepackage{multirow}

\begin{document}
\begin{center}
  {\LARGE Freeze the BCI until the user is ready:\\a pilot study of a BCI inhibitor}\\
  \vspace{0.5cm}
  {\large L. George,  L. Bonnet, A. Lécuyer}\\
  \vspace{0.5cm}
  INRIA, Campus Universitaire de Beaulieu, F-35042 Rennes Cedex, France\\
  \vspace{0.2cm}
  \href{mailto:laurent.f.george@inria.fr}{\{laurent.f.george, laurent.bonnet, anatole.lecuyer\}@inria.fr}
  \vspace{0.8cm}
\end{center}

\begin{abstract}
  \noindent
In this paper we introduce the concept of Brain-Computer Interface (BCI) inhibitor, which is meant to standby the BCI until the user is ready, in order to improve the overall performance and usability of the system. BCI inhibitor can be defined as a system that monitors user's state and inhibits BCI interaction until specific requirements (e.g.\ brain activity pattern, user attention level) are met.
In this pilot study, a hybrid BCI is designed and composed of a classic synchronous BCI system based on motor imagery and a BCI inhibitor. The BCI inhibitor initiates the control period of the BCI when requirements in terms of brain activity are reached (i.e.\ stability in the beta band). Preliminary results with four participants suggest that BCI inhibitor system can improve BCI performance. 
\end{abstract}

\vspace{1cm}
\section{Introduction}
There are several ways to improve performance of Electroencephalography (EEG) based Brain-Computer Interface (BCI) systems: improving classification methods, developing new signal processing algorithms or increasing EEG hardware efficiency are the most common directions taken~\cite{lottereview}. One recent approach intends to combine different paradigms into a hybrid system~\cite{hybridbci}. For example, the Error Related Potential can be detected during a BCI-based interaction to correct the interaction and increase overall performance~\cite{hybriderp}.
 
However relatively few studies try to improve the performance of BCI systems by focusing on interaction techniques and usage. 
An approach is to use a brain switch~\cite{hybridbci, brainswitchmason} which can be described as a preliminary BCI that allows the user to activate the BCI interaction at will. This system allows reducing false positive in asynchronous BCI context as confirmed in~\cite{pfurtb2010}. 

In this paper we propose an implicit and complementary system which we have named BCI inhibitor. The main objective is to increase performance of a BCI by using a paradigm that will activate the BCI system only if the best conditions are met, with respect to the current state of the user. Thus BCI inhibitor will avoid classifying EEG features when the system detects that the current user's state will undoubtedly lead to an erroneous result.
We evaluated our solution through a pilot study involving motor imagery tasks.

\section{Concept}
BCI inhibitor can be defined as a system that pauses the BCI until specific conditions (e.g. optimum condition in terms of user's state) are met. BCI inhibitor is in relation with the recent idea of a brain switch, but in the case of BCI inhibitor it is not the user that will intend to explicitly activate the BCI. Instead the BCI inhibitor monitors user state to assess the readiness of the user. For this reason BCI inhibitor can be viewed as an implicit (or passive) counterpart to the brain-switch. We expect that the combination of a classical BCI with a BCI inhibitor will result in an improvement of the overall performance, along with more comfort for the user. 
Recently, Panicker et al. described a P300 speller BCI enhanced with a constant flickering~\cite{Panicker}. The P300-based BCI was paused when no Steady State Visual Evoked Potential (SSVEP) response was detected. This system can be seen as a BCI inhibitor in which the inhibition condition is ``user is not looking at the screen'' and the inhibition signal is a SSVEP measured by EEG.  Several other sensor channels like electromyography (EMG) or electrooculography (EOG) could be chosen as inhibitor signal. However the monitoring of the brain activity through EEG seems to be particularly adapted and could provide relevant information to the BCI inhibitor system. Numerous EEG markers appear to be useful to inhibit a BCI: Error potentials, rhythms associated to attention level etc. Another approach is to use features directly correlated to the BCI control signal. This will allow inhibiting the system until some specific conditions about the control signal are met.
The usage of BCI inhibitor seems to be relevant to both asynchronous and synchronous BCI. In the asynchronous context, we can for example think of a hybrid BCI consisted of a Brain switch, followed by a BCI inhibitor that ensures that the user is in the desired state before enabling the control BCI that actually drives the interaction. In the context of synchronous BCI an inhibitor can be used between phases to put the system in standby until specific conditions are satisfied. 

To evaluate the inhibitor process based on this last idea, we designed a hybrid BCI by combining a synchronous motor imagery based BCI that uses beta ERS posterior to feet movement (``beta rebound''),  with a BCI inhibitor that checks whether the signal in the beta frequency band does not show any burst of activity before starting the control period.

\section{Method}
\paragraph{Participants:}Four participants took part in the experiment, respectively aged of 27, 24, 28 and 26. They had never used any motor imagery based BCI system  before.
\paragraph{Setup:}EEG signals were recorded using a g.USBAmp (G.Tec) amplifier, sampled at 512 Hz. The setup was made of 7 electrodes, positioned according to the 10-20 system: a ground electrode (located on AFz position) and a reference electrode (located on the left earlobe), along with five measurement electrodes on Cz, C1, C2, FCz and Cpz. This EEG setup allowed us to record the EEG activity related to motor imagery of the feet~\cite{lottevr}. The application is based on the Virtual Reality application ``Use the force'' presented by Lotte et al.~\cite{lottevr}. A virtual spaceship is displayed on the screen. The goal is to lift the spaceship by doing motor-related tasks: real and imaginary feet movements. Whenever a burst is detected in the beta activity related to feet movement, the application raises the spaceship proportionally. Instructions can be displayed, asking the participant to either stand still, move or stop.

\paragraph{Procedure:} The experiment was divided into 2 parts: a baseline and the series of motor imagery trials, using real movements or imaginary movements.
The baseline consisted of a 25 sec period, where the participants were asked to stand still and relaxed, eyes opened. No feedback was provided during the baseline which was done once, at startup. The trial sequence is inspired by the startup sequence of an athletics run: Ready, Steady, Go.
During one trial, the participants were instructed to relax (``Ready'') for a certain period of time, then waited for 1 sec (``Steady''). Finally the ``Move'' instruction is displayed during 3 sec, followed by ``Stop'' during 3 sec. The movement of the feet done during the ``Move'' phase was instructed to be either real or imaginary. The ``Stop'' instructed the user to stop doing movement which should induced a beta rebound. 
The BCI inhibitor was either activated or deactivated during these trials without telling the participants. The ``Ready'' phase lasted 3 sec without inhibitor. Once the inhibitor was activated, the duration of the ``Ready'' phase varied from a minimum of 0.5 sec to a maximum of 10 sec.
Participants were asked to start with 6 real movement sessions, followed by 6 imaginary movement sessions. The BCI inhibitor was activated on half of the sessions, randomized to eliminate an order effect. Each session was made of 10 trials with 4 sec between trials. The whole experiment (setup, trials, and questionnaire) lasted about 1 hour.

\paragraph{Signal Processing:}
EEG acquisition and online processing were conducted using the open-source software OpenViBE~\cite{openvibe}. The EEG signal is band-pass filtered in 2-40 Hz band. Then, a Laplacian spatial filter centered on Cz is computed. The signal was then filtered in the Beta band (16-24Hz). A band power technique was applied to compute the power of the Beta band. 
We distinguish then two signals processed: the Control Signal (CS) that is used to control the spaceship, and the Inhibitor Signal (IS) used by the BCI inhibitor to decide to either launch or not launch the BCI-based interaction.

We define the \textbf{Control Signal (CS)}  as the beta band power extracted on a 1 s window every 100 ms. The last 4 features were averaged with a moving window to produce a smooth control signal. To detect the post movement Beta ERS, the CS was compared to a threshold: $\mathsf{Th1} = {\mathsf{baseline}_{\mathsf{mean}}} + 3 *{\mathsf{baseline}_{\mathsf{std}}}$; where $\mathsf{baseline}_{\mathsf{mean}}$ and $\mathsf{baseline}_{\mathsf{std}}$ correspond respectively to the average and standard deviation of the control signal during the baseline phase. 

We define the \textbf{Inhibitor Signal (IS)} as the beta band power extracted on a 2 s window every~500 ms. The IS was compared to a threshold $\mathsf{Th2} = \mathsf{baseline}_{\mathsf{mean}} + 1*\mathsf{baseline}_{\mathsf{std}}$. If the computed control signal stayed below $\mathsf{Th2}$~99\% of the time then the inhibition was deactivated and the BCI started. The maximum time of the inhibition was 10~s, after this the BCI started anyway.

\section{Results}
Table~\ref{table1} shows the participants' performance for each condition (inhibitor on vs.\ inhibitor off, real movements vs.\ imaginary movements) and the duration of the \emph{``Ready'' phase}. To assess the participants' performance we counted the number of false positives (FP) and the number of true positives (TP). A false positive occurs if the value of CS went at least once above Th1 during a ``Move'' phase. A true positive occurs if CS signal went at least once above Th1 during a ``Stop'' phase. What happened during the other phases was not taken into account.  We also computed the Hit-False (HF) difference which is equal to the number of TP minus the number of FP.
\begin{table}[h]
  \begin{center}
	  \small{
	\caption{Performance achieved with the motor imagery BCI with and without BCI inhibitor in real and imaginary condition and mean duration of the \emph{``Ready'' phase}. Last row provides the average values over participants.\label{table1}} 

  \begin{tabulary}{\textwidth}{|C|CCCCCC|}
	\hline
	      & Task  &   Inhibitor       & Duration of the  & FP & TP & HF\\
		  &   &         & \emph{``Ready Phase'' (sec)} & & & \\
	\hline
	\multirow{4}{*}{Subject 1} & \multirow{2}{*}{Real}  & on & $1.85\pm1.03$ & 9/30     &  29/30 & 20 \\
		&  & off & $3.00\pm 0.00$   &  3/30 & 30/30 & 27 \\
		\cline{2-7}
		&\multirow{2}{*}{Imaginary}& on & $1.30\pm0.63$ & 6/30 & 14/30  & 8\\
		& & off & $3.00\pm0.00$ & 2/30 & 12/30 & 10\\
		\hline
	\multirow{4}{*}{Subject 2}& \multirow{2}{*}{Real}  &  on & $3.89\pm2.94$ & 19/30     &  28/30 & 9 \\
		&  &  off & $3.00\pm0.00$ &  24/30 & 29/30  & 5\\
		\cline{2-7}
		& \multirow{2}{*}{Imaginary} & on &  $1.25\pm0.65$ & 15/30 & 18/30  & 3\\
		& & off & $3.00\pm 0.00$ & 18/30 & 14/30  & -4\\
		\hline
		\multirow{4}{*}{Subject 3}& \multirow{2}{*}{Real}  &  on & $6.56\pm3.26$ & 14/30     &  30/30 &16 \\
		&  &  off & $3.00\pm0.00$ &  21/30 & 30/30 & 9\\
		\cline{2-7}
		&\multirow{2}{*}{Imaginary} & on &  $4.37\pm3.18$ & 8/30 & 17/30 & 9 \\
		& & off & $3.00\pm 0.00$ & 19/30 & 22/30 & 3\\
		\hline
		\multirow{4}{*}{Subject 4}& \multirow{2}{*}{Real}  &  on &  $3.04\pm2.88$ & 8/30     &  30/30 &22 \\
		&  &  off & $3.00\pm0.00$ &  13/30 & 28/30 & 15\\
		\cline{2-7}
		&\multirow{2}{*}{Imaginary} & on &  $2.65\pm2.86$   & 8/30 & 16/30 & 8 \\
		& & off & $3.00\pm 0.00$ & 16/30 & 14/30 & -2\\

		\hline\hline
		\multirow{4}{*}{Average}& \multirow{2}{*}{Real}  &  on & 3.84  & 12.5/30     &  29.25/30 & 16.75 \\
		&  &  off & $3.00$ &  15.25/30 & 29.25/30 & 14.00 \\
		\cline{2-7}
		&\multirow{2}{*}{Imaginary} & on & 2.39 & 9.25/30 & 16.25/30 & 7.0 \\
		& & off & $3.00$ & 13.75/30 & 15.5/30 & 1.75  \\

		\hline
  \end{tabulary}
  }
  \end{center}
\end{table}

\section{Discussion}
Results suggest that BCI inhibitor works and provides an effect on the BCI behavior. It is materialized by different inhibition times for each user. 
Even if we have to take some cautions considering the limited number of participants it seems that the BCI inhibitor is able to improve the system performance: the average Hit-False difference over subjects is higher when the inhibitor was enabled for real movement condition 
(16.75 vs.\ 14.0) and imaginary condition (7.0 vs.\ 1.75). This result is mainly due to the reduction of false positive (e.g. subjects~2, 3 and 4). 

After each session, participants were asked whether they felt a difference between the two conditions (with and without inhibitor) and to quantify it on a scale between 1 (not at all) and 7 (very). Suprisingly the participants reported only a small difference ($2.2\pm0.92$), which suggests that this BCI inhibitor tends to be transparent to the users' perception. 

The participants could also provide additional comments about each condition. Subject~1 was the only one who identified a different timing between the two conditions. During sessions with inhibitor enabled, he reported to be ``caught by surprise'' due to very short inhibition delay. This suggests that setting a higher minimum inhibition's duration could help to avoid such an effect. 

\section{Conclusion}
In this paper we have introduced the concept of BCI inhibitor which can be defined as a system that pauses the BCI until some specific conditions are met. We presented a pilot study in the case of a synchronous motor imagery based BCI. Preliminary results with four participants suggest that inhibition process can be used to improve system performance. These hypotheses should be confirmed with more subjects. 
Future work should also address the use of BCI inhibitor for other paradigms such as P300 and SSVEP\@. Exploring adapted inhibitor signals (e.g. EEG markers correlated to level of attention) seems to be particularly relevant in these cases.

\section*{Acknowledgments}
This work was supported by the French National Research Agency within the OpenViBE2 project (ANR-09-CORD-017). The authors would also like to thank Fabien Lotte(INRIA Bordeaux Sud-Ouest) for its helpful remarks. 

\bibliographystyle{unsrt}
\small{%
\bibliography{bibliography}
}

\end{document}